\documentclass[epj,twocolumn]{webofc}
\usepackage[varg]{txfonts}   
\usepackage{CJKutf8}
\woctitle{LHCP 2013}
\begin{document}
\title{Enhanced $B_d^0 \to \mu^+\mu^-$ Decay: What if?}
%
%

\author{George W.S. Hou\inst{1}\fnsep\thanks{\email{wshou@phys.ntu.edu.tw}} 
}

\institute{Department of Physics, National Taiwan University,
           Taipei, Taiwan 10617
          }

\abstract{%
If the very rare $B_d^0 \to \mu^+\mu^-$ decay is enhanced to
(3--4) $\times 10^{-10}$ level or higher, it can be discovered
with existing 2011-2012 LHC data. It might then cast some doubt
on the Higgs boson nature of the 126 GeV boson, since
the likely explanation would be due to the fourth generation $t^\prime$ quark.
There is a mild motivation from the known tension in $\sin2\beta/\phi_1$.
If discovery is made before the 13 TeV run, then the $b \to d$ quadrangle
 (modulo $m_{t'}$) would suddenly fall into our lap.
Continued pursuit in future runs can probe below $3 \times 10^{-10}$.
}
\maketitle

\section{Introduction: from Straub to Stone}
\label{intro}

Despite the euphoria around the discovery of a
Higgs-like new boson a year ago,
there is a strong sense of accompanying disappointment
 --- No New Physics.
This holds true not only for the energy frontier,
but for the flavor frontier as well.
Although there was great hope, as the mantle
passed from the Tevatron (and the B factories)
to the LHC from 2008 to 2011,
all hints for possible New Physics (NP)
in $b\to s$ transitions evaporated.
This culminated in the LHCb measurement of $\sin\phi_s$ in 2011
that was consistent with Standard Model (SM) expectations.

A parallel, and much watched, saga is the pursuit of $B_q \to \mu^+\mu^-$.
The experimental measurement has progressed tremendously,
as recorded by the series of ``Straub plots''~\cite{Straub} from 2010 to 2012,
together with the backdrop of possible theories.
The drive has been the potentially huge enhancement
by exotic scalar effects inspired by supersymmetry (SUSY),
but such enhancement has now been excluded by the first evidence,
uncovered by the LHCb experiment~\cite{LHCbBsmumu},
which is consistent with SM.

In the Straub plots, most models of enhancement for
$B_d^0 \to \mu^+\mu^-$ have now been eliminated by the SM-like
$B_s^0 \to \mu^+\mu^-$ rate measured by LHCb.
It was noted by Stone~\cite{Stone} in his talk at the ICHEP 2012 conference,
that the experimental bound on $B_d^0 \to \mu^+\mu^-$ is still
an order of magnitude away from SM expectation,
and one ``finger'' sticks out in the Straub plot:
the 4th generation could saturate this bound.
Stone, however, followed standard wisdom and stated:
``The 125 GeV Higgs observation kills off 4th generation models
as the production cross section would be 9 times larger
and decays to $\gamma\gamma$ suppressed'',
hence viewed the possibility as unviable.

We would like to caution, however, that the SM Higgs
is flavor-blind. Flavor people should always keep CKM-extension
in mind, and also since the Higgs boson itself does not enter
these loops.
Furthermore, the 126 GeV boson could still be a ``dilaton''~\cite{dilaton}
(we refer to the Appendix for a discussion).
In short, we urge the experiments to keep on searching,
with gusto!

We stress that, with $B_s^0 \to \mu^+\mu^-$ SM-like
does not preclude the possibility that $B_d^0 \to \mu^+\mu^-$ gets enhanced,
especially with a 4th generation (4G) since
$V_{t'd}^*V_{t'b}$ and $V_{t's}^*V_{t'b}$ are independent parameters.
This may in fact be the ``last chance'' for genuine NP
in the flavor sector at LHC8 (8 TeV LHC).
Now, 4G is purportedly ``killed" by observation
of the 126 GeV boson. But if one makes a discovery of
$B_d^0 \to \mu^+\mu^-$ for LHC7$+$8, the same argument
could be turned around to cast some doubt on the ``SM Higgs''
nature of the 126 GeV boson. The stakes are therefore very high.
To elucidate this in the cultured city of Barcelona, the host city
for the First LHCP Conference, I draw on a well known ``idiom" in
Chinese opera (and the literature through millennia):\begin{CJK}{UTF8}{bkai}
回馬槍,\end{CJK}
or ``Turn horse around and thrust (the spear)''.
The image is a victorious general in hot pursuit
of his defeated counterpart.
The victor closes in, and the situation gets desperate.
Suddenly, the doomed turns his horse around and thrusts ...
Drama!

A discovery of $B_d^0 \to \mu^+\mu^-$ above $4\times 10^{-10}$ or so,
achievable with existing data, could constitute 
such a$\,$\begin{CJK}{UTF8}{bkai}回馬槍,\end{CJK}
and would make things extremely interesting.

There is in fact a mild motivation for anomalous behavior
in $b \to d$ transitions, i.e. the well-known~\cite{Lunghi:2008aa, Buras:2008nn} tension in $\sin2\Phi_{B_d} \equiv \sin2\beta/\phi_1$,
between the directly measured value of
\begin{equation}
\sin2\beta/\phi_1 = 0.679 \pm 0.020,
\end{equation}
and SM expectation
\begin{align}
\sin2\beta/\phi_1 =
\begin{cases}
0.76  & \text{for}~|V_{ub}|^{\rm ave} \\
0.63  & \text{for}~|V_{ub}|^{\rm excl},
\end{cases}
\end{align}
inferred via $\beta/\phi_1 \cong \arg \lambda_t^{\rm SM}$.
Let us elaborate. Within SM,
\begin{equation}
\lambda_t^{\rm SM} = -\lambda_u -\lambda_c
 \simeq -|V_{ud}||V_{ub}|e^{-i\phi_3} + |V_{cd}||V_{cb}|,
\end{equation}
with $\lambda_i \equiv V_{id}^*V_{ib}$,
and the right-hand side are all directly measurable.
Taking standard PDG values of $|V_{ud}| = 0.974$,  $|V_{cd}| = 0.23$
and $|V_{cb}| = 0.041$, as well as
$\gamma/\phi_3 = (68^{+11}_{-10})^\circ$
(LHCb will dominate in next round of PDG),
we find Eq.~(2), which reflects
a different tension in the measured values of $|V_{ub}| \simeq$
$4.41 \times 10^{-3}$ and $3.23 \times 10^{-3}$ respectively,
extracted via inclusive or exclusive semileptonic $B$ decays,
with a mean value of $4.15 \times 10^{-3}$.
For either value, there is tension between Eqs.~(1) and (2).

The 4G $t'$ quark can easily alleviate this tension
by a new CKM factor $\lambda_{t'} = V_{t'd}^*V_{t'b}$,
\begin{equation}
\lambda_t = \lambda_t^{\rm SM} -\lambda_{t'},
\end{equation}
where $\lambda_t^{\rm SM}$ is given in Eq.~(3),
as the $b\to d$ triangle becomes a quadrangle
\begin{equation}
\lambda_u + \lambda_c + \lambda_t + \lambda_{t'} = 0.
\end{equation}
We parameterize
\begin{equation}
\lambda_{t'} = r_{db}\,e^{i\phi_{db}}.
\end{equation}
where we maintain the phase convention that
$\lambda_c = V_{cd}^*V_{cb}$ is practically real,
while $\lambda_u = V_{ud}^*V_{ub}$ is basically the same as in SM.
We turn now to constrain $r_{db}$ and $\phi_{db}$.

\section{Constraints and Formulas}
\label{constraints}

We first note that $b\to d\gamma$ processes are
hard to separate from $b\to s\gamma$ processes. Furthermore,
they are hard for LHCb to measure, and in any case
they are insensitive to virtual $t'$ effects.
The $B\to \pi\pi$ decay modes have been studied in
detail, but they suffer from hadronic effects (they
are certainly not better than $B\to K\pi$).
Thus, besides our target $B_d\to \mu^+\mu^-$ rate,
the main constraints are the $B_d$ mixing parameters
$\sin2\Phi_{B_d}$ (i.e. $\sin2\beta/\phi_1$) and $\Delta m_{B_d}$.
We shall also employ $B^+\to \pi^+\mu^+\mu^-$,
which was measured~\cite{LHCb:2012de} by LHCb only a year ago,
the rarest measured $B$ meson decay to date.

The formulas for $B_d$ mixing are well known,
\begin{align}
&\Delta m_{B_d} 
 \simeq \frac{G_F^2M_W^2}{6\pi^2}m_{B_d}
        \hat B_{B_d}f_{B_d}^2 \eta_B
        \left|\Delta_{12}^d\right|, \nonumber \\
&\sin 2\Phi_{B_d} 
 \simeq \sin \left(\arg \Delta_{12}^d\right),
\end{align}
where the short distance $t$ and $t'$ box functions are
\begin{align}
\Delta_{12}^d \equiv&\ (\lambda_t^{\rm SM})^2 S_0(x_t) 
  + 2\lambda_t^{\rm SM}\lambda_{t^\prime} \Delta S_0^{(1)}
  + \lambda_{t^\prime}^2 \Delta S_0^{(2)}, \\
\Delta S_0^{(1)} \equiv&\ \tilde S_0(x_t,x_{t^\prime}) - S_0(x_t),\\
\Delta S_0^{(2)} \equiv&\ S_0(x_{t^\prime}) -2 \tilde S_0(x_t,x_{t^\prime}) +S_0(x_t),
\end{align}
and $x_i = m_i^2/M_W^2$.
The hadronic uncertainty is mainly in
\begin{equation}
f_{B_d}\hat B_{B_d}^{1/2} = (227 \pm 19)\ {\rm MeV}.
\end{equation}

Our purpose is to illustrate the parameter space
where the current bound of
\begin{equation}
{\cal B}(B_d \to \mu^+\mu^-) < 8.1\times 10^{-10},
\end{equation}
can be saturated by 4G effect. The SM-nature allows
us to use the ``Buras ratio'', i.e. normalizing by
the branching ratio by $\Delta m_{B_d}$,
\begin{align}
&\hat{\mathcal B}(B_d \to \mu^+\mu^-)
\equiv \frac{\mathcal B (B_d \to \mu^+\mu^-)}{\Delta m_{B_d}}
 \Delta m_{B_d}^{\rm exp} \nonumber \\
&= C \frac{\tau_{B_d}\Delta m_{B_d}^{\rm exp}}{\hat B_{B_d}}\frac{\eta_Y^2}{\eta_B}
 \frac{ \left| \lambda_t^{\rm SM} Y_0(x_t) +\lambda_{t^\prime}\Delta Y_0 \right|^2}
{\left| \Delta_{12}^d \right|}
\end{align}
where $\Delta Y_0 = Y_0(x_{t'}) - Y_0(x_t)$.
The ratio eliminates the hadronic parameter $f_{B_d}$,
and one is left with the milder uncertainty in
bag parameter $\hat B_{B_d}$.
For SM, the $\lambda_t^{\rm SM}$ factor also cancels,
and one recovers SM result of $1.1 \times 10^{-10}$,
and there is little sensitivity to $|V_{ub}|$.

The recent measurement of $B^+\to \pi^+\mu^+\mu^-$ by LHCb
\begin{align}
{\cal B}(B^+\to \pi^+\mu^+\mu^-)
= (2.3\pm 0.6 \pm 0.1) \times 10^{-8},
\end{align}
is the first observation of a $b\to d\ell^+\ell^-$ mode,
corresponding to $\sim 25$ events with 1 fb$^{-1}$ data.
The result is consistent with SM expectations.

To interpret the LHCb result, not only is there form factor dependence,
tt turns out that the technology for computing this
decay is not yet fully developed.
To have better numerical control, we follow Ref.~\cite{Beneke:2001at, Beneke:2004dp}
and take Wilson coefficients at NLO, but LO amplitudes in QCDF.
We then integrate in the $q^2$ region of $(1,\ 6)$ GeV$^2$,
and take the ratio of 4G vs SM result,
\begin{align}
R_{\pi\mu\mu} \equiv
\frac{\mathcal B(B^+\to \pi^+\mu^+\mu^-)|_{\rm 4G}}
{\mathcal B(B^+\to \pi^+\mu^+\mu^-)|_{\rm SM}}.
\end{align}
Our ``ansatz'' then is to plot $R_{\pi\mu\mu}$ contours,
assuming that if it exceeds 2 to 3, then likely LHCb
would not have claimed consistency with SM.
This is clearly not as good as the zero crossing point
$q_0^2$ for $A_{\rm FB}(B\to K^*\mu\mu)$,
but this is the first observation of rare $b\to d\ell\ell$ decays,
compared to the decade-long exploration of $b\to s\ell\ell$ processes.

\begin{figure*}[t!]
\centering
{\includegraphics[width=67mm]{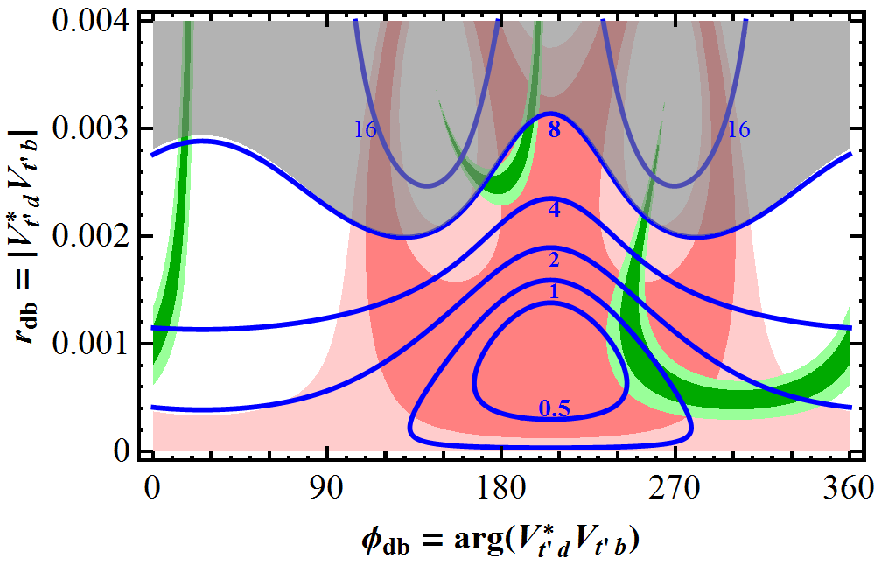}
 \includegraphics[width=67mm]{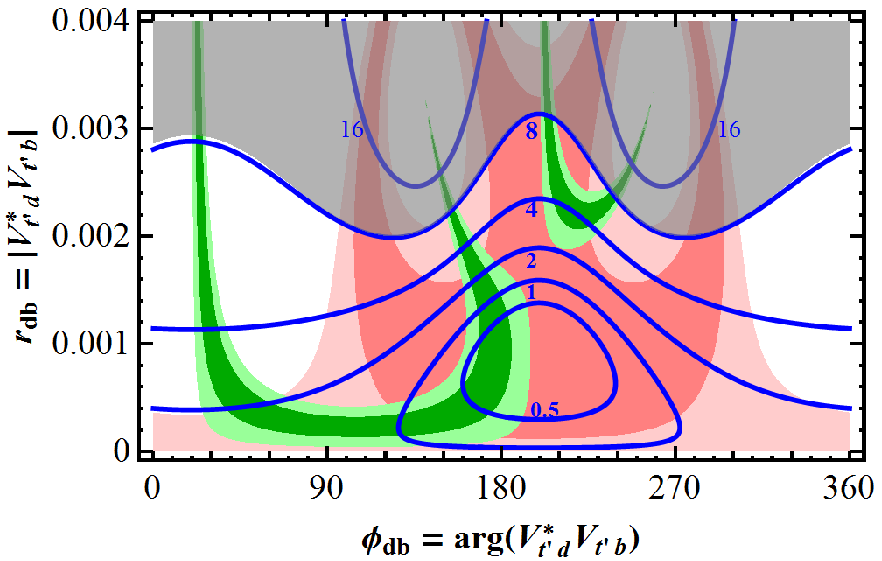}
}
\caption{
Allowed regions in $r_{db}$--$\phi_{db}$ plane for $m_{t'}= 700$ GeV
and average (left) and exclusive (right) $|V_{ub}|$ values.
Solid-blue lines are labeled $10^{10}{\cal B}(B_d \to \mu^+\mu^-)$ contours,
where the shaded region is excluded by LHC experiments.
The narrow green-shaded contours correspond to
1 and 2$\sigma$ regions of $\sin2\Phi_{B_d}$,
while the broad pink-shaded contours correspond to
regions of $\Delta m_{B_d}$ allowed by Eq.~(11).
} \label{DeltamBd-700}
\end{figure*}

\begin{figure*}[t!]
\centering
{\includegraphics[width=67mm]{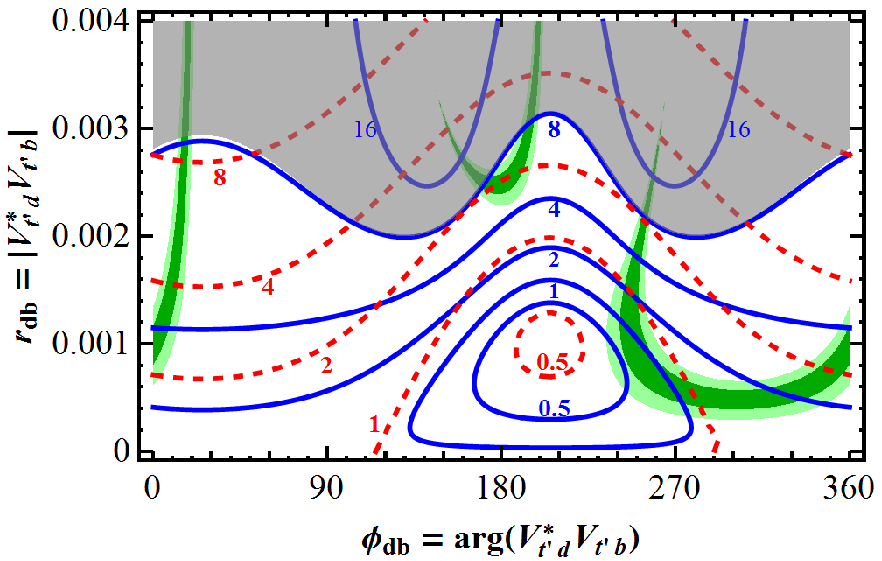}
 \includegraphics[width=67mm]{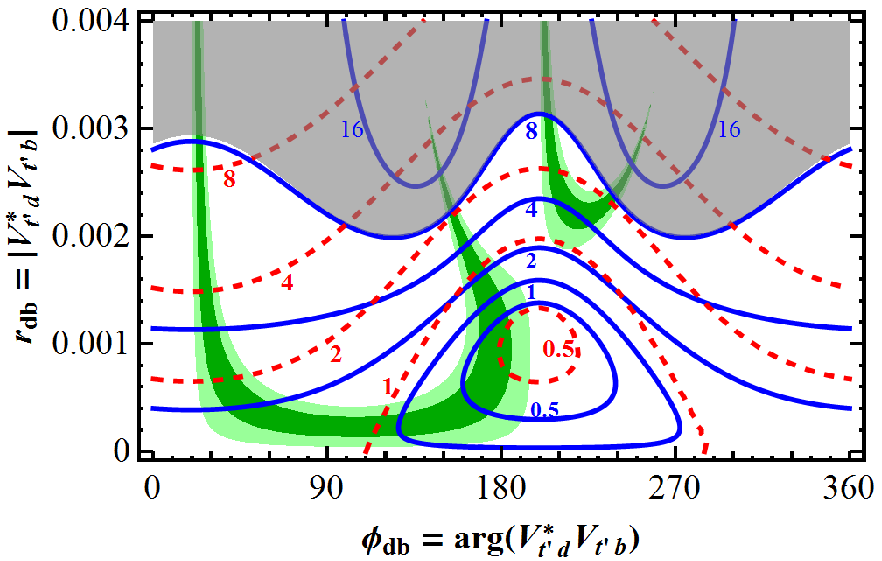}
}
\caption{
Same as Fig.~1, but with $\Delta m_{B_d}$ replaced by
contours (red-dashed) of $R_{\pi\mu\mu}$ of Eq.~(15),
which should be less than 2 to 3.
} \label{B2pimumu-700}
\end{figure*}

\section{\boldmath Phenomenological Results for Heavy $t'$}
\label{results}

Current bounds on 4G quark masses are rather stringent,
so we illustrate with $m_{t'} = 700$ GeV.
The narrow green shaded bands in Fig.~1 are the 1 and 2$\sigma$ ranges
in the $r_{db}$--$\phi_{db}$ plane for
$\sin2\Phi_{B_d}$ allowed by experimental measurement of Eq.~(1).
The left and right panels are for taking
$|V_{ub}| = 4.15 \times 10^{-3}$ and $3.23 \times 10^{-3}$, respectively, for
the mean (between inclusive and exclusive) and exclusive values
from semileptonic $B$ decay.
Similarly, the broad pink shaded region
is for $\Delta m_{B_d}$ allowed by lattice error in Eq.~(11).
The solid-blue lines are labeled contours of 0.5, 1, 2, 4, 8 and 10
for $10^{10}{\cal B}(B_d \to \mu^+\mu^-)$, and
the bound on ${\cal B}(B_d \to \mu^+\mu^-)$ according to Eq.~(12)
is displayed as the gray exclusion region.

Consider the left panel of Fig.~1, which is for
$|V_{ub}| = 4.15 \times 10^{-3}$ from
the mean of inclusive and exclusive values
from semileptonic $B$ decay.
The $\Delta m_{B_d}$-allowed region (pink) is generally
much more accommodating than $\sin2\Phi_{B_d}$ (green)
because of the associated hadronic uncertainty.
However, the green band around $\phi_{db} \simeq 0$
allowed by $\sin2\Phi_{B_d}$ is excluded by $\Delta m_{B_d}$,
because the $t'$ effect would add constructively with the top effect.
The bound on ${\cal B}(B_d \to \mu^+\mu^-)$ (gray)
limits the strength of $r_{db} = |V_{t'd}^*V_{t'b}|$.
However, around $(r_{db},\ \phi_{db}) \sim (0.0025,\ 180^\circ)$
and $(0.002,\ 252^\circ)$, ${\cal B}(B_d \to\mu^+\mu^-)$
could reach above $4 \times 10^{-10}$,
i.e. enhanced by 4 times over SM.
We relatively arbitrarily set this as
the discovery zone for 2011-2012 LHC data,
and call the above two regions A and B, respectively.
Extending below region B, ${\cal B}(B_d \to\mu^+\mu^-)$
quickly drops to $2 \times 10^{-10}$ as
$r_{db}$ drops from 0.002 to 0.0015,
and below $r_{db} \simeq 1$, there is a
crescent-shaped band allowed by $\sin2\Phi_{B_d}$ and $\Delta m_{B_d}$,
where ${\cal B}(B_d \to\mu^+\mu^-)$ is close to SM expectation,
and the phase $\phi_{db}$ varies over a broad range,
from 235$^\circ$ to 330$^\circ$.
We call this region C.

For the right panel, where one takes $|V_{ub}|$ from
exclusive semileptonic $B$ decay, one roughly switches
the sign on $\phi_{db}$ from the left panel, and
we call the respective allowed regions A$'$, B$'$ and C$'$.
A point each will be taken from A and A$'$ for later illustration.

Now we compare the implications of the bound of Eq.~(14),
$B^+\to \pi^+\mu^+\mu^-$.
We elaborated in the previous section that
we take the ratio $R_{\pi\mu\mu}$ of Eq.~(15) to
eliminate form factor dependence,
but the branching fractions for both 4G and SM
are integrated over only the range of $q^2 \in (1, 6)$ GeV$^2$,
for sake of numerical control. Our ansatz then is to
disallow $R_{\pi\mu\mu}$ greater than 2--3.
Comparing Fig. 1 and 2, we see that $\Delta m_{B_d}$ is
slightly more powerful than ${\cal B}(B^+\to \pi^+\mu^+\mu^-)$
in excluding the $\sin2\Phi_{B_d}$-allowed branch near $\phi_{db} \sim 0$.
But otherwise, the 3 allowed regions of Fig.~1 survive
the $R_{\pi\mu\mu} \lesssim 2$--3 criteria,
which provides a useful sanity check.

We have illustrated how $B_d \to \mu^+\mu^-$ can be
enhanced up to the current bound by 4G.
For heavier $t'$, say $m_{t'} = 1000$ GeV which is far beyond unitarity bound,
we find that $B_d \to \mu^+\mu^-$ can be
more easily enhanced up to the current bound,
with $|V_{t'd}^*V_{t'b}|$ dropping by more than 1/2,
but region A (large $r_{db}$ at $\phi_{db} \sim 180^\circ$)
gets eliminated.

\begin{figure*}[t!]
\centering
{\includegraphics[width=130mm]{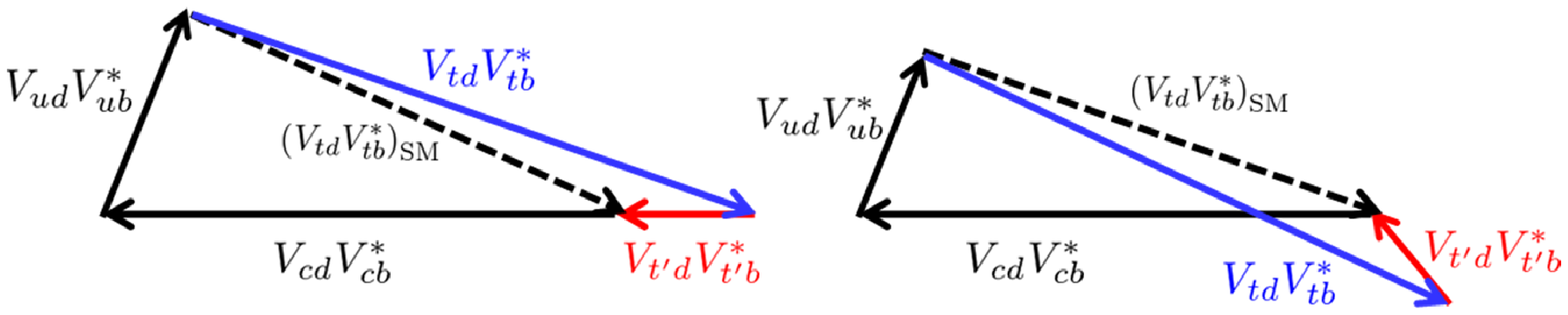}
}
\caption{
Sample $b\to d$ quadrangles
for $\lambda_{t^\prime}=V_{t^\prime d}^*V_{t^\prime b}=0.0025\, e^{i 180^\circ}$
 (left), and
$0.0023\, e^{i 230^\circ}$
 (right).
} \label{fig:BdQuad}
\end{figure*}

\section{Discussion}
\label{discussion}

Eq.~(5) gives a quadrangle, 
rather than the familiar triangle of SM3.
If $B_d \to \mu^+\mu^-$ gets enhanced hence discovered in the near future,
with the implied finite values for $V_{t'd}^*V_{t'b}$,
we would all of a sudden have the $b\to d$ quadrangle in our lap
(modulo $m_{t'}$).

Let us take two examples for $m_{t'} = 700$ GeV.
We take
\begin{equation}
\lambda_{t'} = V_{t'd}^*V_{t'b} = 0.0025\, e^{i 180^\circ}, \quad
0.0023\, e^{i 230^\circ}.
\end{equation}
which are from region A of Fig.~1(a)
and region A$'$ of Fig.~1(b), corresponding to
average $|V_{ub}| = 4.15 \times 10^{-3}$ and
exclusive $|V_{ub}| = 3.23 \times 10^{-3}$, respectively.
The two quadrangles are shown in Fig.~\ref{fig:BdQuad},
in usual PDG convention.

The quadrangles of Fig.~\ref{fig:BdQuad} can be
easily constructed from Eq.~(5), with the help of Eqs.~(3) and (4).
From Eq.~(3), $\lambda_u$ and $\lambda_c$ remain basically unchanged
from SM3 case,
hence form a triangle with $\lambda_t^{\rm SM}$,
which is the dashed vector in Fig.~\ref{fig:BdQuad}.
Eq.~(4) then prescribes how to form the full
$\lambda_t$ with $\lambda_{t'}$ as given in Eq.~(16),
and as illustrated in Fig.~\ref{fig:BdQuad}.
This figure also illustrates how the
$\sin2\Phi_{B_d}$ tension is generated, or accounted for,
since the SM triangle is constructed from
tree level measurements assuming CKM3 unitarity.
We are reminded by these quadrangles that
4G can in principle provide sufficient~\cite{Hou:2008xd} CPV for
generating the Baryon Asymmetry of the Universe
(BAU), which is a strong motivation for 4G.

Although CPV for BAU does not depend strongly on the
actual size of the quadrangle (but depend rather on
the strength of $m_{b'}$ and $m_{t'}$),
we remark that the CKM4 product $|V_{t'd}V_{t'b}|$
values of Eq.~(16) are quite large.
This is because $|V_{t'b}| < 0.1$ is necessary~\cite{Hou:2010mm} to
satisfy constraints,
hence the values of $|V_{t'd}|$ are a few \%,
larger than $|V_{td}|$ in SM.
If we learned anything from the $\sin2\Phi_{B_s}$ measurement,
it is that the CKM hierarchy seem upheld, even in the case that 4G exists.
If we maintain the CKM hierarchy, that $|V_{t'd}|$ 
should at least not be larger than $|V_{td}|$,
then though ${\cal }(B_d \to \mu^+\mu^-)$ enhanced
by a factor of 4 beyond SM is possible, 
it \emph{does not seem particularly likely}.
Heavier $t'$ masses, such as the 1000 TeV case, lead to a sharper
drop in $|V_{t'd}V_{t'b}|$, hence tends towards
more ``naturally small'' CKM4 factors,
although the quadrangle would become harder to draw or picturise.

In the end, the strength of $|V_{t'd}V_{t'b}|$ is
an experimental question, where the 
$B_d \to \mu^+\mu^-$ mode can provide decisive input, soon.

\section{Conclusion}
\label{conclusion}

2013 is a pivotal year: 
If the $B_d \to \mu^+\mu^-$ rate is enhanced over SM
by a factor of 4 or more, we will discover it!
It is certainly in experimental range (mainly LHCb and CMS),
and there is some motivation from $\sin2\Phi_{B_d}$ ``anomaly''.
If discovery is made with 2011--2012 data, then
it could give a boost to 4G quarks, with hopes again
for CPV-4-BAU. 
In return, a discovery could cast some doubt on the ``Higgs'' nature
of the 126 GeV boson --- could ``it'' be itself from New Physics?
Of course, theorists would be scrambling to account
for $B_d \to \mu^+\mu^-$, but much fine-tuning would be needed,
compared with the natural CKM4 parameter provided by 4G.

It would be a great impetus to particle physics if
discovery is made soon.
What is more likely, however, as we have experienced in the past 2--3 years,
is that the bound would be pushed down towards SM once again.
Even so, we would still need to push the search for 
the $B_d \to \mu^+\mu^-$ mode in the 13-14 TeV runs at the LHC.\\

\noindent\textbf{Acknowledgement.}\
We thank M. Kohda and F. Xu for collaboration.

%
%
%

\end{document}